\documentclass[aps,twocolumn,pra,superscriptaddress,showpacs,tightenlines]{revtex4}
\usepackage{epsfig,graphicx,times}
\usepackage{float}
\usepackage{amstext}
\usepackage{amsmath}
\usepackage{amssymb}
\usepackage{graphicx}
\usepackage{latexsym}
\usepackage{bm}
\usepackage[colorlinks,citecolor=blue, linkcolor=blue,hyperindex,CJKbookmarks,dvipdfm]{hyperref}

\begin{document}

\title{%Control optical fields output from
Controllable optical output fields from an optomechanical system with a mechanical driving}
\author{Xun-Wei Xu}
\affiliation{Beijing Computational Science Research Center, Beijing 100084, China}
\author{Yong Li}
\email{liyong@csrc.ac.cn}
\affiliation{Beijing Computational Science Research Center, Beijing 100084, China}
\affiliation{Synergetic Innovation Center of Quantum Information and Quantum Physics,
University of Science and Technology of China, Hefei, Anhui 230026, China}
\date{\today }

\begin{abstract}
%  We investigate the properties of the optical fields output from a cavity optomechanical system where the cavity is driven by a strong coupling and a weak probe optical fields and the mechanical resonator is driven by a coherent mechanical pump. When the frequency of the mechanical pump matches the frequency difference between the coupling and probe optical fields, there exist two optical components at the frequency of the probe field, one is induced directly by the weak probe optical field and the other one is generated by the strong coupling optical field and the mechanical driving field. The quantum  destructive or instructive  interference between these two optical components provides us a way to control the properties of the optical field output from the optomechanical system at the frequency of the probe field. We demonstrate that the large positive or negative group delay of the output field at the frequency of probe field can be achieved and tuned by adjusting the phase and amplitude of the mechanical driving field. Moreover, the optical FWM field generated from the coupling and probe fields, can interfere with the field generated from the coupling field and mechanical pump. The strength of the output field at the frequency of FWM field can also be controlled (enhanced and suppressed) by tuning the phase and amplitude of the mechanical driving field. We show that the power of the output field at the frequency of FWM field can be suppressed to zero or enhanced so much that it can be comparable with and even larger than the power of the input probe optical field.

  We investigate the properties of the optical output fields from a cavity optomechanical system, where the cavity is driven by a strong coupling and a weak probe optical fields and the mechanical resonator is driven by a coherent mechanical pump. When the frequency of the mechanical pump matches the frequency difference between the coupling and probe optical fields, due to the interference between the different optical components at the same frequency, we demonstrate that the large positive or negative group delay of the output field at the frequency of probe field can be achieved and tuned by adjusting the phase and amplitude of the mechanical driving field. Moreover, the strength of the output field at the frequency of optical four-wave-mixing (FWM) field also can be controlled (enhanced and suppressed) by tuning the phase and amplitude of the mechanical pump. We show that the power of the output field at the frequency of the optical FWM field can be suppressed to zero or enhanced so much that it can be comparable with and even larger than the power of the input probe optical field.
\end{abstract}

\maketitle

%42.50.Wk   Mechanical effects of light on material media, microstructures and particles (see also 87.80.Cc Optical trapping in biology and medicine)
%42.50.Nn	Quantum optical phenomena in absorbing, amplifying, dispersive and conducting media; cooperative phenomena in quantum optical systems
%42.65.Dr	Stimulated Raman scattering; CARS (for Raman lasers, see 42.55.Ye)
%42.65.Ky	Frequency conversion; harmonic generation, including higher-order harmonic generation (see also 42.79.Nv Optical frequency converters)

\section{Introduction}

The field of optomechanical system that a movable mirror couples to a high-quality cavity via radiation pressure force has drawn much research attention in the past several years~\cite%
{KippenbergSci08,MarquardtPhy09,AspelmeyerPT12,AspelmeyerRMF14}. One of the
motivations is that optomechanical system is an important candidate for
signal transmission and storage. In analog to the optical responses in
atomic system, a form of induced transparency called optomechanically
induced transparency (OMIT) was pointed out theoretically~\cite{AgarwalPRA10}. Soon afterwards, the phenomenon of OMIT as well as optomechanically induced absorption and amplification was reported for observing in various optomechanical systems~\cite{WeisSci10,TeufelNat11,Safavi-NaeiniNat11,MasselNat11,MasselNC12,HockeNJP12,ZhouNP13,KaruzaPRA13,SinghNN14,FongArx14}. One important result concomitant with the modification of the optical
response is a dramatic variation of the phase of the transmitted field and
this will lead to a positive or negative group delay of the propagating
field. Based on the OMIT, optomechanically induced absorption and
amplification, optomechanical systems may be used for storage of optical
signals~\cite{JiangEPL11,ChenPRA11,ZhanJPB13,WangPRA14,JingArx14}, and the slowing, advancing and switch of optical pulses have been demonstrated experimentally~%
\cite{Safavi-NaeiniNat11,ZhouNP13}.

Moreover, if an optomechanical cavity is driven by a strong
coupling field with frequency $\omega_{c}$ and a weak probe field with frequency
$\omega_{p}$, then, due to the optical radiation pressure, besides the fields at the applied field frequencies $\omega_{c}$ and $\omega_{p}$, a optical four-wave-mixing (FWM) field with frequency $2\omega_{c}-\omega_{p}$ will be generated in the system. The FWM field induced by the radiation pressure in optomechanical systems has been investigated theoretically~\cite{HuangPRA10,JiangJOSAB12}, and it was shown that the strength of the FWM field can be enhanced by increasing the pump power of the strong coupling field. However, usually the power of the output FWM field is still much less than the power of the input probe field, and to our knowledge there is no experimental report on the demonstration of FWM field in optomechanical systems to date.

Recently, the optical response properties in an optomechanical system with
the mechanical resonator driven by an additional mechanical field was
considered theoretically~\cite{JiaArx14}. It was found in Ref.~\cite{JiaArx14} that the optomechanically induced transparency, amplification or absorption for the probe field can be controlled by adjusting the phase and amplitude of the
strong coupling field. In experiments, by applying both optical and mechanical driving fields to a multimode-cavity optomechanical system with both the optomechanical coupling and parametric phonon-phonon coupling~\cite{FanArx14}, a cascaded optical transparency was observed and the extended optical delay and higher transmission as well as optical advancing were demonstrated.

Based on the optomechanical system with a mechanical pump~\cite{JiaArx14,FanArx14}, in this paper we investigate in detail how to control the properties of the optical output fields from the cavity by adjusting the phase and amplitude of the mechanical pump. In the optomechanical system under consideration, the cavity is driven by a strong coupling optical field with requency $\omega_{c}$ and a weak probe optical field with frequency $\omega_{p}$, the mechanical resonator is driven by a coherent mechanical field with frequency $\omega_{q}$. Due to the parametric coupling between the optical and mechanical modes, there are two kinds of optical components in the output field of the optomechanical system: the one induced by the strong coupling and weak probe optical fields with frequencies $\omega_{c}\pm m \delta$ ($\delta \equiv\omega _{p}-\omega _{c}$) and the one generated by the strong coupling optical field and the mechanical driving field (sidebands) with frequencies $\omega_{c}\pm n\omega_{q}$, where $m$ and $n$ are integers~\cite{MarquardtPRL06,SchliesserNP08,XiongPRA12}.
In particular, when the resonant condition $\omega_{q}=\delta \equiv \omega_{p}-\omega_{c}$ is fulfilled, the destructive or instructive quantum interference between these two kinds of optical output components provides us a way to control the properties of the optical fields output from an optomechanical system. For example, the optical delay of the output field at the frequency of optical probe field with frequency $\omega_{p}=\omega_{c}+\omega_{q}$ can be tuned accordingly by adjusting the phase and amplitude of the mechanical driving field. Moreover, the strength of the output field at the frequency of the FWM field $2\omega_{c} - \omega _{p}=\omega_{c}-\omega_{q}$ can also be controlled (i.e., enhanced and suppressed) by tuning the phase and amplitude of the mechanical driving field.
%due to the quantum interference at the  resonant condition. Moreover,
%The controllable of the output optical field is due to the quantum interference between the two kinds of optical fields in the optomechanical system, one is induced by the strong coupling and weak probe optical fields and the other one is generated by the strong coupling optical field and the mechanical driving field.

This paper is organized as follows: In Sec.~II, we introduce the theoretical model of the optomechanical system with the driving of a strong coupling optical field, a weak probe optical field, and a mechanical pump. In Sec.~III, the properties of the sidebands induced by parametric coupling in this optomechanical system is discussed. The effect of the mechanical driving field on the group delay of the output field at the frequency of probe optical field is investigated in Sec.~IV. In Sec.~V, the output field at the frequency of the FWM field with the presence of mechanical driving field is studied. We draw our conclusions in Sec.~VI.

\section{Theoretical model}

\begin{figure}[tbp]
\includegraphics[bb=4 380 592 594, width=8.5 cm, clip]{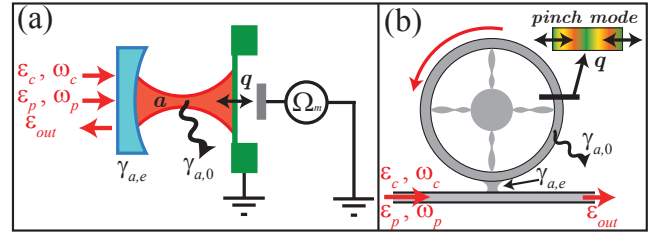}
\caption{(Color online) (a) Schematic of the optomechanical system for a
Fabry-Perot optical cavity with a moving mirror (mechanical resonator). The
optical cavity is driven by a coupling field and a probe field from the
left-hand side. A mechanical pump $\Omega_{m}$ is applied to the mechanical
resonator via an ac voltage source. (b) Experimental setup of the micro-wheel
optomechanical system for optical mode coupling to a mechanical pinch mode
through radiation pressure force, where the mechanical pump is realized by
piezoelectric drive with radio frequency signal~\cite{FanArx14}.}
\label{fig1}
\end{figure}

We consider an optomechanical system consisting of a single-sided optical
cavity with a moving mirror (mechanical resonator) in the right-hand side
[Fig.~\ref{fig1}(a)]. The optical cavity is driven by a strong coupling field
(amplitude $\varepsilon _{c}$, frequency $\omega _{c}$) and a weak probe
field (amplitude $\varepsilon _{p}$, frequency $\omega _{p}$) from the left-hand side. A mechanical pump of strength $\Omega_{m}=2\varepsilon
_{m}\cos \left( \omega _{q}t+\phi _{m}\right) $ (amplitude $\varepsilon _{m}$, frequency $\omega _{q}$, phase $\phi _{m}$) is applied to the mechanical resonator via a tunable ac voltage source. In the rotating frame at the
frequency of the optical coupling field $\omega _{c}$, the Hamiltonian of
the system is given by
\begin{eqnarray}  \label{Eq1}
H/\hbar &=&\Delta _{a}a^{\dag }a+\frac{1}{2}\omega _{m}\left(
q^{2}+p^{2}\right) +ga^{\dag }aq  \notag \\
&&+i\left( \varepsilon _{c}a^{\dag }+\varepsilon _{p}e^{-i\delta t}e^{-i\phi
_{p}}a^{\dag }- \mathrm{H.c.}\right)  \notag \\
&&+2q\varepsilon _{m}\cos \left( \omega _{q}t+\phi _{m}\right) ,
\end{eqnarray}%
where $a$ ($a^{\dag }$) denotes the bosonic annihilation (creation) operator
of the cavity mode with resonance frequency $\omega _{a}$ in the absence of
moving mirror; $q$ and $p$ are the dimensionless displacement and momentum
operators of the moving mirror with resonance frequency $\omega _{m}$. $%
\Delta _{a}=\omega _{a}-\omega _{c}$ ($\delta =\omega _{p}-\omega _{c}$) is
the frequency detuning between the cavity mode (probe field) and the
coupling field. We assume that $\varepsilon _{c}$ and $\varepsilon _{p}$ are real, and $\phi
_{p}$ is the phase difference between the probe and
coupling fields. This Hamiltonian (\ref{Eq1}) can also be realized by a
micro-wheel cavity with optical mode coupling to a mechanical pinch mode
through radiation pressure force [Fig.~\ref{fig1}(b)] and the mechanical
pinch mode is piezoelectrically driven by a radio frequency signal as in Ref.~\cite%
{FanArx14}.

%In this paper, we will deal with the problem that how to control the properties of the output optical fields from an optomechanical system by a mechanical driving field in the presence of a strong coupling optical field and a weak probe optical field.
By the Heisenberg equations and the factorization assumption like $\left\langle qa\right\rangle =\left\langle q\right\rangle \left\langle a\right\rangle $, one gets the mean value equations for the operators of
the optical and mechanical modes
\begin{eqnarray}
\frac{d}{dt}\left\langle a\right\rangle &=&-\left[ \gamma _{a}+i\left(
\Delta _{a}+g\left\langle q\right\rangle \right) \right] \left\langle
a\right\rangle  \notag  \label{Eq2} \\
&&+\varepsilon _{c}+\varepsilon _{p}e^{-i\delta t}e^{-i\phi _{p}},
\end{eqnarray}%
\begin{eqnarray}
&&\frac{d^{2}}{dt^{2}}\left\langle q\right\rangle +\gamma _{m}\frac{d}{dt}%
\left\langle q\right\rangle +\omega _{m}^{2}\left\langle q\right\rangle
\notag  \label{Eq3} \\
&=&-\omega _{m}g\left\langle a^{\dag }\right\rangle \left\langle
a\right\rangle -2\omega _{m}\varepsilon _{m}\cos \left( \omega _{q}t+\phi
_{m}\right) ,
\end{eqnarray}%
where $\gamma _{m}$ is the mechanical damping rate; $\gamma _{a}=\gamma
_{a,0}+\gamma _{a,e}$ is the cavity damping rate, including the intrinsic
loss $\gamma _{a,0}$ and the damping through the left-side mirror $\gamma
_{a,e}$.

Firstly, we examine the mean values of the operators in the steady state and
denote $\alpha _{0}$ and $q_{0}$ the (zeroth-order) steady-state mean values for the cavity field and
mechanical displacement when $\varepsilon _{p}\rightarrow 0$ and $\varepsilon_{m}\rightarrow 0$. The
mean values in the steady state can be expressed as
\begin{equation}  \label{Eq4}
\alpha _{0}=\frac{\varepsilon _{c}}{\gamma _{a}+i\Delta },
\end{equation}
\begin{equation}  \label{Eq5}
q_{0}=-\frac{g}{\omega _{m}}\left\vert \alpha _{0}\right\vert ^{2},
\end{equation}
where $\Delta =\Delta _{a}+gq_{0}$ is the effective detuning.

Generally, the mean values can be written as the sum of the steady state mean values and the time
dependent ones, i.e., $\left\langle a\right\rangle =\alpha _{0}+\widetilde{%
\alpha }\left( t\right) $ and $\left\langle q\right\rangle =q_{0}+\widetilde{%
q}\left( t\right) $.
Since the probe optical field and mechanical driving field are much weaker than the strong coupling optical field, i.e., $\left\{ \left\vert\varepsilon _{p}\right\vert ,\left\vert \varepsilon _{m}\right\vert \right\} \ll \left\vert \varepsilon _{c}\right\vert
$, we have $|\alpha _{0}|\gg |\widetilde{\alpha }\left(t\right) |$ and $|q_{0}|\gg |\widetilde{q}\left( t\right) |$, then Eqs.~(\ref{Eq2}) and (\ref{Eq3}) can be linearized by keeping the linear terms of the small time-dependent values,
\begin{equation}
\frac{d}{dt}\widetilde{\alpha }=-\left( \gamma _{a}+i\Delta \right)
\widetilde{\alpha }-iG\widetilde{q}+\varepsilon _{p}e^{-i\delta t}e^{-i\phi
_{p}},  \label{Eq6}
\end{equation}%
\begin{eqnarray}
&&\frac{d^{2}}{dt^{2}}\widetilde{q}+\gamma _{m}\frac{d}{dt}\widetilde{q}%
+\omega _{m}^{2}\widetilde{q}  \notag   \\
&=&-\omega _{m}\left[ \varepsilon _{m}e^{i\left( \omega _{q}t+\phi
_{m}\right) }+G\widetilde{\alpha }^{\ast }+\mathrm{c.c.}\right] , \label{Eq7}
\end{eqnarray}%
where $G=g\alpha _{0}$ is the effective optomechanical coupling rate.

With using the ansatz:
\begin{eqnarray}
%$
\widetilde{\alpha }&=&\alpha _{p+}e^{-i\delta t}+\alpha
_{p-}e^{i\delta t}+\alpha _{m+}e^{-i\omega _{q}t}+\alpha _{m-}e^{i\omega
_{q}t}, \label{alpha} \\ %$, $
\widetilde{q}&=&q_{p}e^{-i\delta t}+q_{p}^{\ast }e^{i\delta
t}+q_{m}e^{-i\omega _{q}t}+q_{m}^{\ast }e^{i\omega _{q}t}, %$,
\end{eqnarray}
one can obtain the solution of the coefficients as
\begin{eqnarray}
\alpha _{p+} &=&\frac{\varepsilon _{p}e^{-i\phi _{p}}}{\gamma _{a}+i\left(
\Delta -\delta \right) +i\chi \left( \delta \right) \left\vert G\right\vert
^{2}}, \label{Eq8}\\
\alpha _{p-} &=&\frac{-iG}{\gamma _{a}+i\left( \Delta +\delta \right) }%
q_{p}^{\ast }, \label{Eq9}\\
q_{p} &=&\frac{\chi \left( \delta \right) G^{\ast }\varepsilon _{p}e^{-i\phi
_{p}}}{\gamma _{a}+i\left( \Delta -\delta \right) +i\chi \left( \delta
\right) \left\vert G\right\vert ^{2}},\label{Eq10}
\end{eqnarray}%
\begin{eqnarray}
\alpha _{m+} &=&\frac{-i\chi \left( \omega _{q}\right) G}{\gamma
_{a}+i\left( \Delta -\omega _{q}\right) +i\chi \left( \omega _{q}\right)
\left\vert G\right\vert ^{2}}\varepsilon _{m}e^{-i\phi _{m}}, \label{Eq11}\\
\alpha _{m-} &=&\frac{-iG}{\left[ \gamma _{a}+i\left( \Delta +\omega
_{q}\right) \right] }q_{m}^{\ast }, \label{Eq12}\\
q_{m} &=&\frac{\chi \left( \omega _{q}\right) \left[ \gamma _{a}+i\left(
\Delta -\omega _{q}\right) \right] }{\gamma _{a}+i\left( \Delta -\omega
_{q}\right) +i\chi \left( \omega _{q}\right) \left\vert G\right\vert ^{2}}%
\varepsilon _{m}e^{-i\phi _{m}},\label{Eq13}
\end{eqnarray}%
where the mechanical effective susceptibility%
\begin{equation}\label{Eq14}
\chi \left( \omega \right) =\frac{-\omega _{m}}{\left( \omega
_{m}^{2}-\omega ^{2}-i\gamma _{m}\omega +\frac{i\omega _{m}\left\vert
G\right\vert ^{2}}{\gamma _{a}-i\left( \Delta +\omega \right) }\right) }.
\end{equation}

Note that in the right hand of Eq.~(\ref{alpha}), the first term means the optical probe field with frequency $\omega_p$, the second terms represents the FWM field with frequency $2\omega_c-\omega_p\equiv\omega_p-2\delta$, the third and fourth terms stand respectively for the (first) upper and lower sidebands generated by the strong coupling optical field and the mechanical driving field with frequencies $\omega_c \pm \omega_q$.

Using the input-output relation~\cite{GardinerPRA85}, the optical field output from the optomehcanical cavity is
given by
\begin{equation}\label{Eq15}
\varepsilon _{\mathrm{out}}=2\gamma _{a,e}\left\langle a\right\rangle
-\varepsilon _{c}-\varepsilon _{p}e^{-i\delta t}e^{-i\phi _{p}}.
\end{equation}%
The output field at the frequency of the probe field ($\omega _{p}$) is
obtained as (normalized to the input power of the optical probe field)%
\begin{equation}
t_{p}=\frac{2\gamma _{a,e}\alpha _{p+}-\varepsilon _{p}e^{-i\phi _{p}}}{%
\varepsilon _{p}e^{-i\phi _{p}}},  \label{Eq16}
\end{equation}%
and the FWM field (frequency $2\omega _{c}-\omega _{p}$) in the
output of the cavity is given by%
\begin{equation}
t_{f}=\frac{2\gamma _{a,e}\alpha _{p-}}{\varepsilon _{p}e^{-i\phi _{p}}}.
\label{Eq17}
\end{equation}%
Besides, due to the mechanical driving field, there are two sidebands ($t_{u}$ for the upper sideband $\omega _{c}+\omega _{q}$ and $t_{l}$ for the
lower sideband $\omega _{c}-\omega _{q}$) generated in the output
field with the amplitudes
\begin{eqnarray}
t_{u}&=&\frac{2\gamma _{a,e}\alpha _{m+}}{\varepsilon _{m}e^{-i\phi _{m}}} \label{Eq18}, \\
t_{l}&=&\frac{2\gamma _{a,e}\alpha _{m-}}{\varepsilon _{m}e^{-i\phi _{m}}}, \label{Eq19}
\end{eqnarray}%
which are normalized to the power of the mechanical driving field.

If the frequency of the mechanical driving field is equal to the frequency detuning between the probe and coupling fields, i.e., $\omega_{q}=\delta\equiv\omega _{p}-\omega _{c}$,
then the output field at the frequency of the probe field is obtained as (normalized to $\varepsilon _{p}e^{-i\phi _{p}}$)
\begin{equation}\label{Eq20}
t_{pu}=t_{p}+\eta t_{u} e^{i\phi },
\end{equation}%
where $\eta =\varepsilon _{m}/\varepsilon _{p}$, $\phi =\phi _{p}-\phi _{m}$. The group delay $\tau _{\mathrm{g}}$ is defined by~\cite{WeisSci10,Safavi-NaeiniNat11,ZhouNP13}
\begin{equation}
\tau _{\mathrm{g}}=\frac{d\theta }{d\delta },  \label{Eq21}
\end{equation}%
where $\theta =\mathrm{arg}\left( t_{pu}\right) $ is the phase dispersion of
the output field at the frequency $\omega _{p}=\omega _{c}+\omega _{q}$. The
definition allows for a negative group delay ($\tau _{g}<0$) which
corresponds to the advancing of the probe field.

At the same resonant condition, the optical FWM field (with frequency $2\omega _{c}-\omega _{p}$) and the lower sideband (with frequency $\omega _{c}-\omega _{q}$) have the same frequency, then the output field at the frequency of the FWM field ($2\omega _{c}-\omega _{p}=\omega _{c}-\omega _{q}$) is given by (normalized to $\varepsilon _{p}e^{-i\phi _{p}}$)
\begin{equation} \label{Eq22}
t_{fl}=t_{f}+\eta t_{l} e^{i\phi }.
\end{equation}%

From Eqs.~(\ref{Eq20}) and (\ref{Eq22}), the output fields are the coherent addition of the two parts (from the optical probe and the mechanical pump, $t_{p}$ and $\eta t_{u}$ or $t_{f}$ and $\eta t_{l} $), and the relative amplitude $\eta$ and phase difference between them $\phi$ can be used to control the properties of the output fields.

In the following sections, we will investigate the properties of the optical fields output from the optomechanical system numerically by the equations obtained in this section.
In the numerical calculations, we use the parameters from a recent experiment~\cite{FanArx14}: $\omega_{m}=2\pi \times 1.094$\,GHz, $\gamma_{a}=2\pi \times 0.255$\,GHz, $\gamma_{a,e}=0.2\gamma_{a}$ and $\gamma_{m}=2\pi \times 16.8$\,kHz. Without loss of generality, $G$ is assumed real and positive in the following discussions. Here and afterwards, we fix $\Delta=\omega_{m}$, which provides the optomechanical cooling for the mechanical resonator and is widely applied in the previous works on OMIT~\cite{AgarwalPRA10,WeisSci10,TeufelNat11,Safavi-NaeiniNat11,MasselNat11,MasselNC12,HockeNJP12,ZhouNP13,KaruzaPRA13,SinghNN14,FongArx14}.

% which can be realized by appropriately choosing the phase of the coupling optical field

\section{Mechanical pump induced sidebands}

In this section, we will discuss about the sidebands induced by the mechanical pump given by Eqs.~(\ref{Eq18}) and (\ref{Eq19}). Here, we only consider the first sidebands (upper sideband with the frequency $\omega _{c}+\omega _{q}$ and lower sideband with the frequency $\omega _{c}-\omega _{q}$) by taking the linear approximation in the derivation of Eqs.~(\ref{Eq18}) and (\ref{Eq19}). The effect of higher-order sidebands can also be discussed by considering the nonlinear terms as in Ref.~\cite{XiongPRA12}.

\begin{figure}[tbp]
\includegraphics[bb=45 236 558 617, width=8.5 cm, clip]{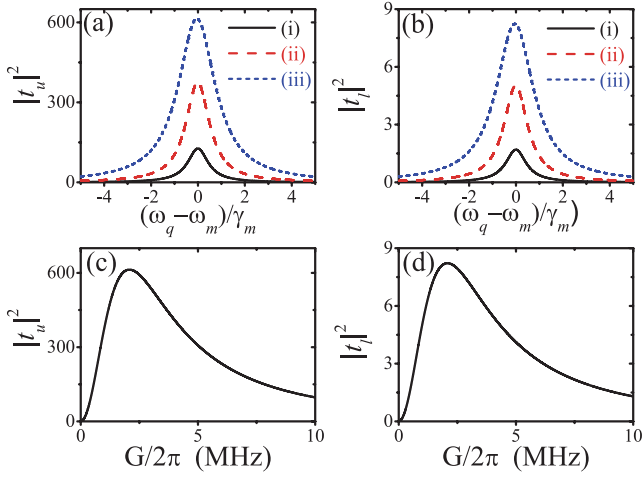}
\caption{(Color online) The strength of the output field at the upper sideband $|t_{u}|^2$ [(a) and (c)] and
lower sideband $|t_{l}|^2$ [(b) and (d)] as functions of the detuning between mechanical mode and mechanical driving field [(a) and (b)] for optomechanical coupling rate taking different values [(i) $G=2\pi\times 0.5$ MHz; (ii) $G=2\pi\times 1.0$ MHz; (iii) $G=2\pi\times 2.0$ MHz], and as functions of the optomechanical coupling rate [(c) and (d)] for mechanical resonator driven resonantly ($\omega _{q}=\omega _{m}$).  Note that we fix $\delta=\omega _{q}$ and $\Delta=\protect\omega_{m}$ here and afterwards. The other parameters are $\omega_{m}=2\pi \times 1.094$ GHz, $\protect\gamma_{a}=2\protect\pi \times 0.255$ GHz, $\protect\gamma_{a,e}=0.2\protect\gamma_{a}$ and $\protect\gamma_{m}=2\protect\pi \times
16.8$ kHz.}
\label{fig2}
\end{figure}

Figure~\ref{fig2} shows the strength of the output fields at the first-sideband ($|t_{u}|^2$ and $|t_{l}|^2$) as functions of the detuning $(\omega _{q}-\omega _{m})/\gamma_{m}$ [panels (a) and (b)] and the optomechanical coupling rate $G/2\pi$ [panels (c) and (d)]. There is a peak for the power spectrum of the output field at the first-sideband around the point $\omega _{q}=\omega _{m}$, and the linewidth of the spectrum is broadened with the increase of the optomechanical coupling rate. Moreover, the strength of the output field at the first-sideband is much stronger than the mechanical driving field ($|t_{u}|^2>1$ and $|t_{l}|^2>1$) around the resonate point $\omega _{q}=\omega _{m}$, and the strength of the output field at the first-sideband is dependent on the strength of the optomechanical coupling rate.

As shown in Fig.~\ref{fig2}(c) and~\ref{fig2}(d), the strength of the output field at the first-sideband increases fast with the optomechanical coupling rate and reaches the maximum around $G= 2\pi\times 2.1$ MHz, which is consistent with the analytical result of the optimal value $G_{\rm{om}}\simeq \sqrt{\gamma_{a}\gamma_{m}}=2\pi\times 2.1$ MHz from Eqs.~(11) and (18). As the incease of $G$ above the optimal value $G_{\rm{om}}$, the strength of the output field at the first-sideband decreases slowly with the optomechanical coupling rate because of the fast broadening of the spectrum when $G$ becomes large.

\section{Optical delay and advancing}

As have been mentioned in the introduction, the optical response
properties in an optomechanical system by coherently driving the mechanical
resonator were investigated in Ref.~\cite{JiaArx14}, and optomechanically
induced transparency, absorption and amplification are predicted for the
probe field by adjusting the phase and amplitude of the coupling optical
field. In addition to that, group delay is another important parameter
to describe the optical responses. In this section, with Eqs.~(\ref{Eq20}) and (\ref{Eq21}), we will show that the optical positive or negative group delay of the output field at the frequency of optical probe field can be controlled by adjusting the phase and amplitude of the mechanical driving field.

\begin{figure}[tbp]
\includegraphics[bb=60 255 549 620, width=8.5 cm, clip]{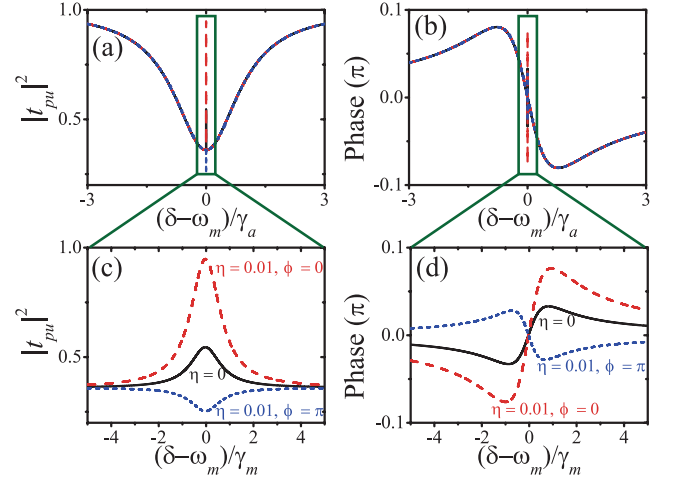}
\caption{(Color online) The output spectrum $|t_{pu}|^2$ [(a) and (c)] and
phase [(b) and (d)] as functions of the detuning between probe and coupling
fields. The black solid line is given for $\protect\eta = 0$; the red dash
line is given for $\protect\eta = 0.01$ and $\protect\phi = 0$ ; the blue
short dash line is given for $\protect\eta = 0.01$ and $\protect\phi =
\protect\pi$. Here $G=2 \pi \times 1.5$ MHz. For the other parameters, see Fig.~\ref{fig2}.}
\label{fig3}
\end{figure}

\begin{figure}[tbp]
\includegraphics[bb=69 369 544 558, width=8.5 cm, clip]{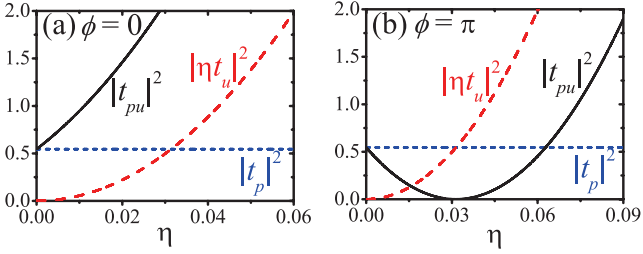}
\caption{(Color online) $|t_{p}|^{2}$, $|\eta t_{u}|^{2}$ and $|t_{pu}|^{2}$ plotted as functions of the normalized amplitude of the mechanical pump $\eta$ for for $G=2 \protect\pi \times 1.5$ MHz with (a) $\phi=0$ and (b) $\phi=\pi$. Here $\protect\delta=\protect\omega_{m} $ and the other parameters are the same as in Fig.~\protect\ref{fig2}.}
\label{fig4}
\end{figure}

The output spectrum $|t_{pu}|^2$ as a function of
the detuning between the probe and coupling fields is presented in Fig.~\ref{fig3}(a) and (c). In the case without the mechanical pump ($\eta=0$), there is a low
efficient transparency window with transmission rate $|t_{pu}|^{2} \approx
0.54$ at $\delta = \omega_{m}$. When the mechanical pump is applied to the
mechanical resonator with $\eta = 0.01$, the transparency of the probe field
will be enhanced with transmission rate $|t_{pu}|^{2} \approx 0.95$ for $\phi=0$ or suppressed with transmission rate $|t_{pu}|^{2} \approx 0.25$ for $\phi=\pi$. This is similar to that reported in Ref.~\cite{JiaArx14}. In order to reveal more about the origin for the enhancing and suppressing of the output field induced by the mechanical pump, we show $|t_{p}|^{2}$, $|\eta t_{u}|^{2}$, and $|t_{pu}|^{2}$ as functions of the normalized amplitude of the mechanical pump $\eta$ in Fig.~\ref{fig4}. When $\phi=0$, there is constructive interference between $t_{p}$ and $\eta t_{u}$, and one can find $|t_{pu}|^{2}\simeq 4|t_{p}|^{2}\simeq 4|\eta t_{u}|^{2}$ around $\eta=0.031$ as shown in Fig.~\ref{fig4}(a). When $\phi=\pi$, there is destructive interference between $t_{p}$ and $\eta t_{u}$, and one can find the strongest destructive interference $|t_{pu}|^{2}\simeq 0$ around $\eta=0.031$ as shown in Fig.~\ref{fig4}(b).

\begin{figure}[tbp]
\includegraphics[bb=49 253 539 620, width=8.5 cm, clip]{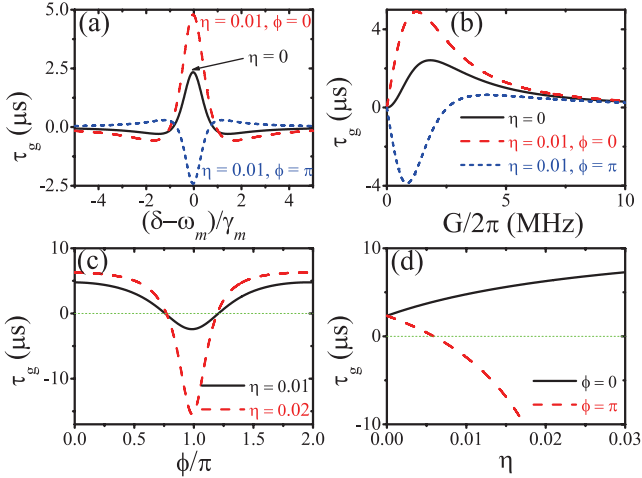}
\caption{(Color online) Group delay $\protect\tau_{g}$ as a function of (a)
the detuning between coupling and probe fields $(\protect\delta-\protect%
\omega_{m})/\protect\gamma_{m}$, (b) the effective optomechanical coupling
rate $G$, (c) phase difference $\protect\phi/\protect\pi$ and (d) strength
ratio $\protect\eta$. In panels (b), (c) and (d), the detuning $\protect\delta=
\protect\omega_{m}$. The other parameters are the same as in Fig.~\protect
\ref{fig2}.}
\label{fig5}
\end{figure}

The phase dispersion and the group delay of the output field at the frequency of the probe field are
shown in Fig.~\ref{fig3}(d) and Fig.~\ref{fig5}(a), respectively. The
enhancement of the optical transparency leads to faster variation of the
phase [see the red dash line in Fig.~\ref{fig3}(d)] and therefore larger optical
group delay of the output field [see the red dash line in Fig.~\ref{fig5}(a)]. Conversely, the suppression of the transparency induces negative derivative of the phase with respect to the frequency [see the blue
short dash line in Fig.~\ref{fig3}(d)] and negative group delay, that is the
advancing of the output field [blue short dash line in Fig.~\ref{fig5}(a)].

The effects of the mechanical pump on the group delay of the output field at the frequency of the probe field are shown in Fig.~\ref{fig5}. From Fig.~\ref{fig5}(a) and (b), the
delay can be increased by the mechanical pump ($\eta = 0.01$) with phase $\phi=0$ or tuned to negative value when $\phi=\pi$. The group delay as a
function of the amplitude $\eta$ of the mechanical pump and phase $\phi$ is
given in Fig.~\ref{fig5}(c) and (d), and the tunable group
delay (positive and negative) of the output field at the frequency of the probe field can be realized efficiently by adjusting the amplitude and phase of the mechanical pump.

\begin{figure}[tbp]
\includegraphics[bb=60 239 532 587, width=4.15 cm, clip]{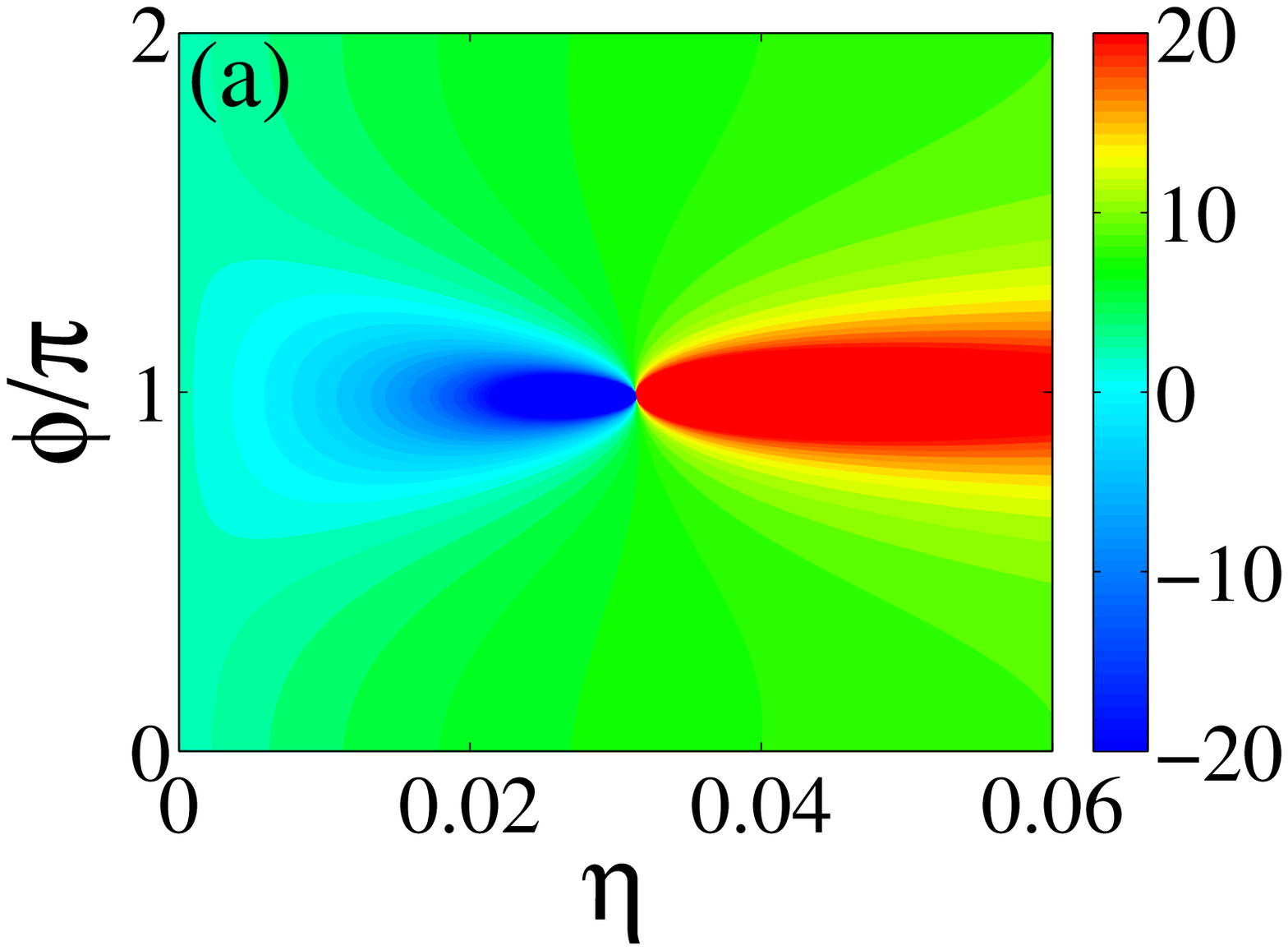} %
\includegraphics[bb=4 10 389 282, width=4.35 cm, clip]{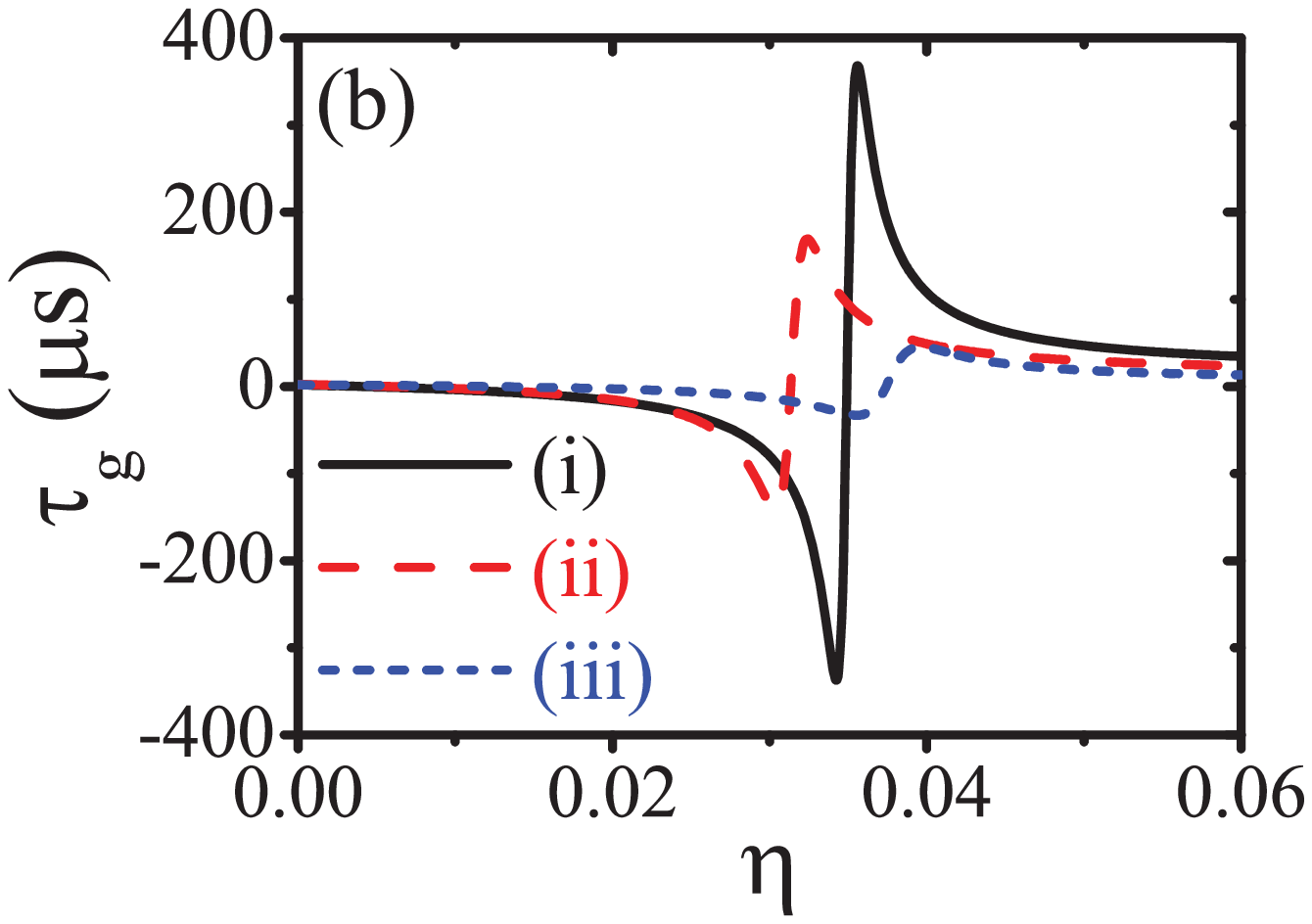}
\caption{(Color online) (a) Group delay $\tau_{\mathrm{g}}$ (in units of $\mu$s) plotted
as a function of $\protect\eta$ and $\protect\phi/\protect\pi$ for $G=2%
\protect\pi \times 1.5$ MHz. (b) $\protect\tau_{\mathrm{g}}$ plotted as a
function of $\protect\eta$ for different values of $G$ [(i) $G=2\pi\times 1.0$ MHz; (ii) $G=2\pi\times 1.5$ MHz; (iii) $G=2\pi\times 3.0$ MHz] with $\protect\phi=%
\protect\pi$. In this figure, the detuning $\protect\delta=\protect\omega%
_{m} $ and the other parameters are the same as in Fig.~\protect\ref{fig2}.}
\label{fig6}
\end{figure}

The group delay as a function of the amplitude and phase of the mechanical
pump is shown in Fig.~\ref{fig6}(a). This plot shows that the group delay
varies fast around the point $\eta \simeq 0.031$ and $\phi=\pi$, i.e., it can be
tuned from the strong negative delay for $\eta $ a little smaller than $0.031$ to the strong positive delay for $\eta $ a little larger than $0.031$ at fixed $\phi=\pi$. This fast variation of group delay induced by the mechanical pump is associated with the changing from the absorption to the transparency of the probe field in the transmission spectrum as shown in Fig.~\ref{fig4}(b). In addition, the group delay as a function of $\eta$ for different values of the effective optomechanical coupling rate $G$ is shown in Fig.~\ref{fig6}(b). Much longer maximal possible delay (advancing) can be obtained for smaller $G$.

\section{Control the output field at frequency of FWM field}

\begin{figure}[tbp]
\includegraphics[bb=46 375 552 571, width=8.5 cm, clip]{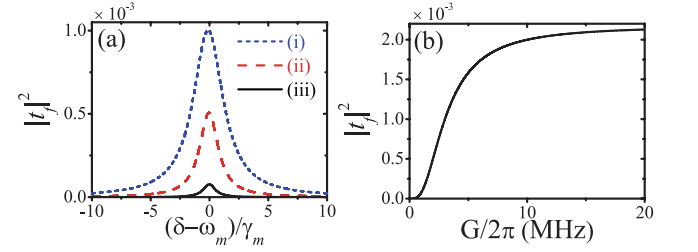}
\caption{(Color online) In the absence of mechanical pump $\protect\eta=0$,
the normalized strength of the output FWM field $|t_{f}|^{2}$ plotted (a) as a
function of the detuning $(\protect\delta-\protect\omega_{m})/\protect%
\gamma_{m}$ for different values of $G$ [(i) $G=2\pi\times 3$ MHz; (ii) $G=2\pi\times 2$ MHz; (iii) $G=2\pi\times 1$ MHz], and (b) as a function of $G$ with $\protect\delta=\protect\omega_{m}$. The other parameters are the same as in Fig.~\protect\ref{fig2}.}
\label{fig7}
\end{figure}

The FWM field in standard optomechanical system has already been studied in
some previous literatures~\cite{HuangPRA10,JiangJOSAB12}. It was shown that
the signal of FWM can be enhanced and exhibits normal-mode splitting as the
increasing of the power of the coupling field (or the effective
optomechanical coupling rate $G$)~\cite{HuangPRA10,JiangJOSAB12}. The output
FWM spectra in the absence of mechanical pump is shown in Fig.~\ref{fig7}.
There are two features for the FWM enhancement by the increase of $G$~\cite%
{HuangPRA10,JiangJOSAB12}: (i) the linewidth of the FWM spectra is broadened
with the strengthening of the FWM field; (ii) the strength of the FWM field
will reach a saturation point [$|t_{f}|^{2}_{\rm sat} \approx (\gamma_{a,e}/\omega_{m})^{2}$] by the increase of $G$ and the saturation strength is much smaller than the input probe field (about $2.2\times 10^{-3}$ for the parameters given in Fig.~\ref{fig7}). In this section, with Eq.~(\ref{Eq22}), we will show that the strength of the output field at the frequency of the FWM field also can be
controlled by the mechanical pump.

\begin{figure}[tbp]
\includegraphics[bb=16 204 560 628, width=8.5 cm, clip]{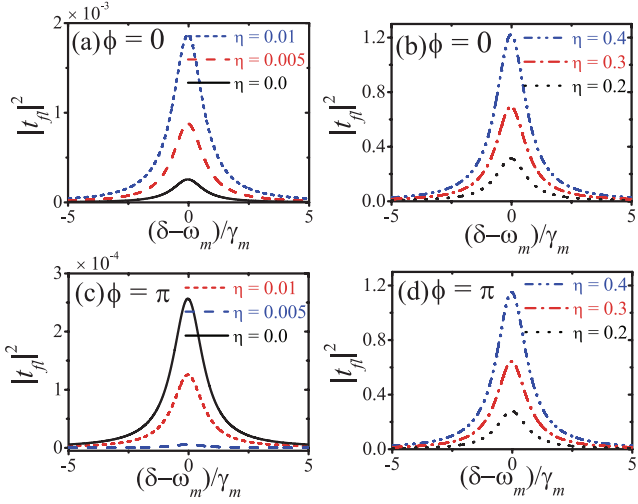}
\caption{(Color online) The normalized strength of the output field at the frequency of the FWM field $%
|t_{fl}|^{2}$ plotted as a function of the detuning $(\protect\delta-\protect%
\omega_{m})/\protect\gamma_{m}$ for different values of $\protect\eta$ with $\phi=0$ in (a) and (b) and $\phi=\pi$ in (c) and (d). The other parameters are the same as in Fig.~\protect\ref{fig2}.}
\label{fig8}
\end{figure}

In Fig.~\ref{fig8}, the normalized strength of output field at the frequency of the FWM field $|t_{fl}|^{2}$ is plotted as a function of the detuning $(\delta-\omega_{m})/%
\gamma_{m}$ for different values of the amplitude of the mechanical pump.
When $\phi=0$, with the strengthening of the mechanical pump, $|t_{fl}|^{2}$ increases significantly [Fig.~\ref{fig8}(a)]. When $\phi=\pi$, with the strengthening of the mechanical pump, $|t_{fl}|^{2}$ are suppressed almost completely for $\eta = 0.005$ as shown in Fig.~\ref{fig8}(c). $|t_{fl}|^{2}$ becomes comparable and even larger than the strength of input probe field with the further increase of $\eta$ for both $\phi=0$ and $\pi$ as shown in Fig.~\ref{fig8}(b) and (d). The linewidth of the output field at the frequency of the FWM field is almost fixed.

\begin{figure}[tbp]
\includegraphics[bb=85 255 516 619, width=8.5 cm, clip]{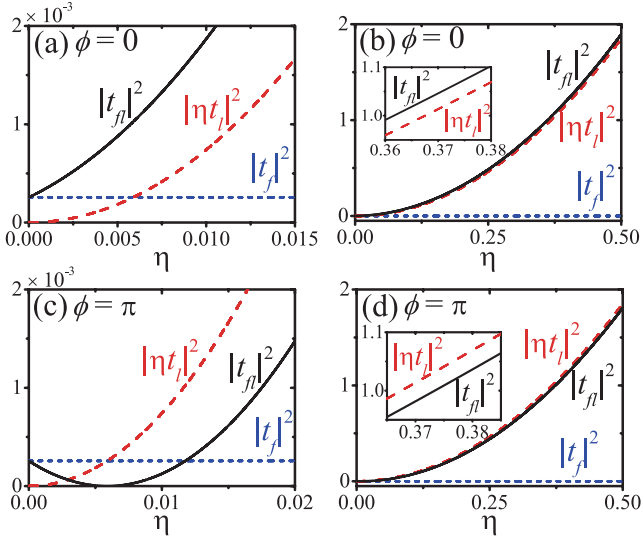}
\caption{(Color online) $|t_{f}|^{2}$, $|\eta t_{l}|^{2}$ and $|t_{fl}|^{2}$ plotted as functions of the normalized amplitude of the mechanical pump $\eta$ with $\phi=0$ in (a) and (b) and $\phi=\pi$ in (c) and (d). Here $\protect\delta=\protect\omega_{m}$ and $G=2 \protect\pi \times 1.5$ MHz. The other
parameters are the same as in Fig.~\protect\ref{fig2}.}
\label{fig9}
\end{figure}

To explain the enhancing and suppressing of the strength of output field at the frequency of the FWM field induced by the mechanical pump, we show $|t_{f}|^{2}|$, $|\eta t_{l}|^{2}$ and $|t_{fl}|^{2}$ as functions of the normalized amplitude of the mechanical pump $\eta$ in Fig.~\ref{fig9}. When $\phi=0$, there is constructive interference between $|t_{f}|^{2}$ and $|\eta t_{l}|^{2}$, e.g., $|t_{fl}|^{2}\simeq 4|t_{f}|^{2}\simeq 4|\eta t_{l}|^{2}$ around $\eta=0.0059$ as shown in Fig.~\ref{fig9}(a). When $\phi=\pi$, there is destructive interference between $|t_{f}|^{2}$ and $|\eta t_{l}|^{2}$, e.g., $|t_{fl}|^{2}\simeq 0$ around $\eta=0.0059$ as shown in Fig.~\ref{fig9}(c). By further strengthening the mechanical pump, the strength of output field at the frequency of the FWM field increases monotonically with no saturation behavior, as shown in Fig.~\ref{fig9}(b) and (d). The normalized strength of the output field at the frequency of the FWM field becomes larger than $1$ when $\eta>0.38$ and this mainly comes from the output field at the lower sideband induced by the mechanical pump.

Analytically, from Eq.~(\ref{Eq22}), the output field at the frequency of the FWM field
will vanish ($t_{fl}=0$) at $\delta=\omega_{m}$ when the amplitude and phase
for the mechanical pump satisfy
\begin{equation}  \label{Eq23}
\eta e^{i\phi }=\frac{-G^{\ast }}{\gamma _{a}+i\left( \Delta -\delta \right)}.
\end{equation}
With the parameters using in Fig.~\ref{fig9}, that is, $G=2\pi \times 1.5$ MHz,
$\Delta=\delta=\omega_{q}=\omega_{m}$, and $\gamma_{a}=2\pi \times
0.255$ GHz, one can find from Eq.~(\ref{Eq23}) that the output field at the frequency of the FWM field
will vanish ($t_{fl}=0$) when $\eta \simeq 0.0059$ and $\phi=\pi$. This
agrees well with the numerical result shown in Fig.~\ref{fig9}(c).

\section{Conclusions}

In summary, we have investigated the properties of the optical output fields from an optomechanical system in the presence of a strong coupling optical field, a weak probe optical field, and a mechanical pump. We demonstrate that the optical delay of the output field at the frequency of optical probe field can be tuned by adjusting the phase and amplitude of the mechanical driving field. This result may have potential applications in quantum information storage and transfer by realizing long optical delay or advancing in optomechanical systems. Moreover, the strength of the output field at the frequency of FWM field can also be controlled (enhanced and suppressed) by tuning the phase and amplitude of the mechanical driving field. The enhancing or suppressing of the output field at the frequency of FWM field is induced by the interference between the FWM field and the lower sideband field and it is worth mentioning that both of them are generated by the nonlinear processes. The efficient enhancing of the output field at the frequency of FWM field by a mechanical pump may provide a route to demonstrating FWM in the optomechanical systems experimentally.

%\section{Acknowledgement}
\vskip 2pc \leftline{\bf Acknowledgement}

We thank Q. Zheng for fruitful discussions. This work is supported by the
Postdoctoral Science Foundation of China (under Grant No. 2014M550019), the
National Natural Science Foundation of China (under Grants No. 11422437, No. 11174027, and No. 11121403) and the National Basic Research Program of China (under Grants No. 2012CB922104 and No. 2014CB921403).

\bibliographystyle{apsrev}
\bibliography{ref}

\end{document}